\input phyzzx
\brokenpenalty=10000

\newbox\figbox
\newdimen\zero  \zero= 0 pt
\newdimen\figmove
\newdimen\figwidth
\newdimen\figheight
\newdimen\textwidth
\newtoks\figtoks
\newcount\figcounta
\newcount\figcountb
\newcount\figlines
\def\figreset{\global\figcounta=-1 \global\figcountb=-1
\global\figmove=\baselineskip
\global\figlines=1 \global\figtoks={ } }
\def\picture#1by#2:#3{\global\setbox\figbox=\vbox{\vskip #1
\hbox{\vbox{\hsize=#2 \noindent #3}}}
\global\setbox\figbox=\vbox{\kern 10pt
\hbox{\kern 10pt \box\figbox \kern 10pt }\kern 10pt}
\global\figwidth=\wd\figbox
\global\figheight=\ht\figbox
\global\textwidth=\hsize
\global\advance\textwidth by - \figwidth }
\def\figtoksappend{\edef\temp##1{\global\figtoks=%
{\the\figtoks ##1}}\temp}
\def\figparmsa#1{\loop \global\advance\figcounta by 1
\ifnum \figcounta < #1
\figtoksappend{ 0pt \the\hsize }
\global\advance\figlines by 1
\repeat }
\def\figparmsb#1{\loop \global\advance\figcountb by 1
\ifnum \figcountb < #1
\figtoksappend{ \the\figwidth \the\textwidth}
\global\advance\figlines by 1
\repeat }
\def\figtext#1:#2:#3{\figreset%
\figparmsa{#1}%
\figparmsb{#2}%
\multiply\figmove by #1%
\global\setbox\figbox=\vbox to 0pt{\vskip \figmove  \hbox{\box\figbox}
\vss }
\parshape=\the\figlines\the\figtoks\the\zero\the\hsize
\noindent
\rlap{\box\figbox} #3}
\def\Buildrel#1\under#2{\mathrel{\mathop{#2}\limits_{#1}}}
\def\llongrarrow{\hbox to 40pt{\rightarrowfill}}

\catcode`@=11 
\newtoks\extranum
\extranum={}
\def\p@bblock{\begingroup \tabskip=\hsize minus \hsize
   \baselineskip=1.5\ht\strutbox \topspace-2\baselineskip
   \halign to\hsize{\strut ##\hfil\tabskip=0pt\crcr
       \the\Pubnum\crcr
       \the\extranum\crcr
       \the\date\cr \the\pubtype\crcr}\endgroup}
\catcode`@=12 
%
%

%

%
%

   \def\frac#1#2{\hbox{${#1\over #2}$}}

\def\ie{{\it i.e.}}    \def\etal{{\it et al.}}

\def\ltsima{$\; \buildrel < \over \sim \;$} \def\simlt{\lower.5ex\hbox{\ltsima}}
\def\gtsima{$\; \buildrel > \over \sim \;$} \def\simgt{\lower.5ex\hbox{\gtsima}}

\def\del{\partial}
\def\pac{Paczy{\'n}ski}
\singlespace
 
\pubnum{} 
\pubtype{} 
\titlepage 
\singlespace

\title {Infimum Microlensing Amplification of the Maximum Number 
        of Images of $n$-point Lens Systems} 
\author{Sun Hong Rhie} 
\address{\IGPP}

\abstract
{The total amplification of a source inside a caustic curve of a binary lens
is no less than 3.  Here we show that the infimum amplification 3 is satisfied
by a family of binary lenses where the source position is at the mid-point
between the lens positions independently of the mass ratio which parameterizes
 the family.  We present a new proof of an underlying constraint that
the total amplification of the two positive images is bigger than that of
the three negative images by one inside a caustic.
We show that a similar constraint holds for an arbitrary class of $n$-point
lens systems for the sources in the `maximal domains'.  We introduce the
notion that a source plane  consists  of {\it graded caustic domains} and the
`maximal domain' is the area of the source plane where a source star results
in the maximum $n^2+1$ images. We show that the infimum
amplification of a three point lens is 7,  and it is 
bigger than $n^2+1-n$ for $n\ge 4$.}

\bigskip

{\it subject headings:} \ gravitational lensing -- planetary systems 
 --  stars: binaries: general 
 
\submit{Astrophysical Journal Letters} 
\endpage


\chapter{Introduction}

Gravitational lensing has been known for three quarters of a century 
and has been extensively used in studies of quasars and clusters of
galaxies. In these cases, the lenses are compound lenses consisting of 
extended objects such as galaxies and clusters of galaxies. The recent
gravitational microlensing experiments looking for baryonic dark matter
(Alcock, \etal, 1993, Aubourg, \etal, 1993, Udalski, \etal, 1993)
study nearby stars ($\lsim 50 $kpc) as  source stars, however,
and the lenses are point lenses.  
They are mostly single lenses but some of them have turned out 
to be binary lenses (Udalski, \etal, 1994, Bennett, \etal, 1995,
Alcock, \etal, 1995, Alard, \etal, 1995).  
The detection rate of binary lenses is expected 
to be about $10\%$ according to Mao and \pac\ (1991).  
One interesting class of binary
lenses is that of planetary systems, and     
there is an effort to look for extra solar earth mass planets using 
gravitational microlensing (Tytler \etal, 1995).   
Obviously, the observational advantage of having to deal with point 
lens systems is that the simplicity allows high precision experiments.
From a theorist's point of view, the current and proposed microlensing 
experiments have brought a necessity to study fine details of the 
cleanest lens systems, which in turn would lead to a more
sophisticated understanding of more uncertain lens systems.

It is well known that when two masses are nearby, they produce a 
gravitational lens with  a rich structure that depends on the 
mass ratio  and separation between them and the lens structure is 
best represented by the caustics.  The caustic curves are closed cuspy 
loops, and neither do they self-intersect nor nest each other.  
\FIG\ternary{The source plane of a symmetric ternary lens with the
seperation (distance between two masses) $l = 1$: The caustics
divide the source plane into a heirachy of ``nested" domains. 
\ The domain marked by $\times$ is  the {\it maximal domain} ${\cal D}^3$, 
and the number of images of a source in the area is 10. 
Incidentally,  $\times$ is at the center of the lens positions.} %
(If $n\ge3$, the lenses  are general enough to feature 
self-intersections and nesting of the caustic loops.  
See figure \ternary\ for an example of a ternary lens.)   
A caustic curve of a binary lens defines an {\it inside}
and an {\it outside} and the number of images of a source is five 
{\it inside} a caustic loop and three {\it outside}. 
Infinity lies in the {\it outside} domain and the number of the
images of a source at infinity is three, one at the source position
and the other two at the lens points.  
Obviously, the source at infinity is
unamplified and the total microlensing amplification $A_{\rm tot} (\infty) =1$.
The amplification of the image at the source position
is one and that of the image at each lens position is zero. 
The amplification of a {\it positive image}
($\equiv$ an image with positive parity)
is always no less than 1, and the total amplification $A_{\rm tot} \ge 1$
in the  {\it outside} domain.  In an {\it inside} domain, a source produces 
two  positive images and hence $A_{\rm tot} > 2$.  In fact, 
$A_{\rm tot} \ge 3$ {\it inside} a caustic as shown recently
by Witt and Mao (1995). Their main result was the following relation.
$$
   \sum_{\rm images} {1\over J} = 1 \ ,  \eqn\Jone
$$
where $J$ is the Jacobian determinant of the lens equation.
In order to demonstrate \Jone, the authors converted the binary lens 
equation inside a caustic into a fifth order polynomial equation of 
$J^{-1}$ using an algebraic computational package {\it Mathematica}.
They also made an empirical suggestion that there is only one  
binary lens configuration that accommodates the infimum amplification 3 
inside a caustic.  

Here we show that there are actually one-parameter family of binary 
lenses where $A_{\rm tot} \ge 3$ is saturated {\it inside} a caustic. 
We also present a simple algebraic proof of \Jone\ and show that 
a similar ``sum rule'' holds in the ``maximal domains" of arbitrary
$n$-point lens systems.  This leads to a necessity to look at the source
planes as consisting of {\it graded caustic domains} $\{{\cal D}^m \}$. 
In section 4, we  discuss the hierarchical structure 
of the source plane of an arbitrary $n$-point lens 
and  infimum microlensing amplifications of 
the sources  in  the ``maximal domains."

\chapter {Inside the Caustics of Binary Lenses}

A lens equation is a mapping from an image position to a source
position on the lens plane transverse to the line of sight. 
If we set the distance scale of the lens plane by the Einstein ring 
radius of the total mass,  the normalized lens equation is given by
(Bourassa, Kantowski and Norton, 1973)
$$
  \omega = z - {\epsilon_1\over (\bar z - x_1)}
             - {\epsilon_2\over (\bar z - x_2)}  \ ,   \eqn\leqtwo
$$
where $z$ and $\omega$ are  the image and source positions in complex
coordinates,  $x_1$ and $x_2$ are the lens positions on the real axis,
and   $\epsilon_1$ and $\epsilon_2$ \ ($\epsilon_1 + \epsilon_2 =1$) are
the fractional masses of the lens elements. 
The Jacobian matrix elements of the lens equation \leqtwo\ are  
$$
  \partial_z\omega =1  \quad ,  \quad
  \partial_{\bar z}\omega = {\epsilon_1\over(\bar z - x_1)^2}
          + {\epsilon_2\over(\bar z - x_2)^2} \  \equiv \bar\kappa \ , 
 \eqn\eqomega
$$
and the Jacobian determinant is 
$J =  |\partial_z \omega|^2  - |\partial_{\bar z}\omega|^2 
   =  1 - |\partial_{\bar z}\omega|^2 \le 1 $.   
Lensing is an inverse mapping of the lens equation
and the inverse Jacobian matrix elements are given by 
$$
 \partial_{\omega} z = {1\over J} \quad , \quad
 \partial_{\bar\omega} z = - {\bar\kappa \over J} \ .
  \eqn\eqomegatwo
$$
For a positive image, $ 0\le J \le 1$ and the amplification 
$A = |J|^{-1} \ge 1$.   
A source inside a caustic has two positive images, and the total
amplification of the positive images   $ A_+ \ge 2$.
Therefore, combined with \Jone, the total amplification $A_{\rm tot} \ge 3$.

In order to find all the lens parameters  where the inequality
$A_{\rm tot} \ge 3$ is saturated, we note that 
the minimum $A_{\rm tot} = 3$ can occur only if both of the positive
images have $J = 1$ (or $\partial_{\bar z}\omega = 0$).
They are the two finite limit points of a binary lens (Rhie, 1995a). 
If $l = |x_1 - x_2|$ denotes the separation between the lens masses,   
$$
  z_{\ast\pm} 
  = x_{\rm acm} \pm i \sqrt{\epsilon_1 \epsilon_2}\,l  \ ; 
  \quad  x_{\rm acm} \equiv \epsilon_1 x_2 + \epsilon_2 x_1 \ .  
  \eqn\bstar
$$
The limit points $z_{\ast\pm}$ are the images of $\omega_{\ast\pm}$  
which we can calculate from the lens equation \leqtwo. If we let
$\epsilon\equiv \epsilon_1 - \epsilon_2$, 
$$
 \omega_{\ast\pm} = x_{\rm acm} 
          - {\epsilon \over l^2}(x_2-x_1) 
     \pm i \sqrt{\epsilon_1\epsilon_2}\,\left(l - {2\over l}\right) \ ,    
  \eqn\eqlimits
$$
and $\omega_{\ast+}$ and $\omega_{\ast-}$ have to be degenerate  
in order to be one source producing two images at the limit points.  
$$
  \omega_{\ast+} = \omega_{\ast-} \quad \Rightarrow \qquad
   l = \sqrt{2}  \quad ; \qquad 
     \omega_{\ast} = {1\over 2} ( x_1 + x_2)  \ .  \eqn\found
$$
In other words, a family of binary lenses with seperation $l =\sqrt{2}$ 
accommodate a source position with the infimum amplification 3 inside
a caustic  
and the source position is at the mid-point of the lens positions 
irrespectively of the mass ratio that parameterizes the family.  

The fact that there are two positive images implies that the source
in \found\ is inside a caustic. (In a binary lensing, the number of
negative images is always bigger than that of positive images by one.) 
The three negative images are on the lens axis and all have $J=-3$.
If the source is at the origin ($x_1 + x_2 = 0$),  the lens 
equation for the negative images (on the real axis) is given by 
$$
   0 = x (x^2 - {3\over 2}) + {\epsilon\over\sqrt{2}} 
     \ ,  \eqn\eqone
$$
where $x_2 > x_1$\  is assumed without loss of generality.
The Jacobian determinant is given by
$$
 J = 1 -  \left({(x - x_{\rm acm})^2
       + \epsilon_1\epsilon_2 l^2\over (x-x_1)^2 (x-x_2)^2} \right)^2
   \equiv 1 - |\kappa|^2 \ .  \eqn\eqJ
$$
When the lens is symmetric ($\epsilon = 0$), the images are at 
$ x = 0 \ , \  \pm\sqrt{3/2}$ and it is easy to verify  
that $J = -3$ for each image.  For  an arbitrary $\epsilon$, 
one can reduce the numerator and denominator of $\,|\kappa|^2\,$ into 
second order polynomials using \eqone\ to find that $|\kappa|^2 = 4$
nd so $J = -3$.  
The family of lenses defined by $l=\sqrt{2}$  have caustics 
consisting of either one connected loop or two. (For a three-loop  
caustic, $l <1$. See  Rhie, 1995a). 
Two-loop caustics occur when the minor mass is small, 
and the source position for the infimum 3 is in the caustic near the minor 
mass.  Inside the caustic (very) near the dominant mass, the amplification
goes to infinity as the mass asymmetry grows because the
caustic approaches that of a single lens.  

Here we demonstrate  that \Jone\ does not hold when the source is  
{\it outside} -- the non-maximal domain of a binary lens -- except when
the source is at infinity.  Let's consider a source located
at the mid-point of a symmetric lens.  
When $l < 2$, the source lies inside the caustic, and there are five 
images:  two off the axis ($z=\pm\sqrt{d^2-1}$; $d = l/2$), 
and the other three ($z=0, \pm\sqrt{d^2+1}$) on the axis.  
As the separation $l$ grows beyond the bifurcation condition
$l=2$,  the caustic splits into two 
closed loops and the source falls in the {\it outside} domain.  Thus, 
two out of the five images should disappear, and  they are the  
two solutions $z=\pm\sqrt{d^2 -1}$, which become real at $l = 2$.    
Both have $J = - 4d^2(d^2-1)\  ( < 0 {\rm for} l > 2 )$.  Therefore,
the other three images have $\sum_{\rm images} J^{-1} > 1$
($ \rightarrow 1$ as $d \rightarrow \infty$).  Of course, the total 
amplification of the three images of a source {\it outside}
can be  dominated by negative images and $\sum_{\rm images} J^{-1} < 1$. 
An example is a source near the off-axis cusps which appear when
the caustic  bifurcates into three loops.

\chapter {$\sum J^{-1} = 1$ for the Maximum Number of Images}

The lens equation \leqtwo\ is a mapping from an image position to  a 
source position: $z, \bar z \mapsto \omega$, and lensing is the inverse
of the lens mapping:  $\omega, \bar\omega \mapsto z$ which is 
either triple-valued or quintuple-valued.  
If we let $z_1 \equiv z - x_1$, $z_2 \equiv z - x_2$ and 
$z_{\circ} \equiv z - x_{\rm acm}$, the binary lens equation looks 
quite simple.
$$
 \omega = z - {\bar z_{\circ}\over \bar z_1 \bar z_2} 
   \equiv z - f(\bar z) \ . \eqn\leqthree
$$
We note that the effect of the masses $\epsilon_j: j=1,2$\  is coupled 
only to the conjugate variable $\bar z$ and the relation between $\omega$ 
and $z$ is universal.   We will see below that
this universality is the underlying reason for the constraint \Jone, 
and thus, it holds not only for binary lenses but also for arbitrary 
$n$-point lenses.

In order to find the inverse function  $z(\omega,\bar\omega)$,
we replace $\bar z$ in \leqthree\  by  $\bar\omega + f(z)$:
$ \omega = z - f(\bar\omega + f(z))$ to find that $z$ satisfies 
a fifth order polynomial
equation,  $0 = g(z;\omega, \bar\omega)$\ (Witt, 1990). 
If $\omega_j; \ j = {\circ}, 1, 2$ is defined  similarly to $z_j$, 
$$
 g(z;\omega,\bar\omega) = (z-\omega)(z_{\circ}+\bar\omega_1 z_1 z_2)
                          (z_{\circ}+\bar\omega_2 z_1 z_2) 
                        - z_1z_2(z_{\circ}+\bar\omega_{\circ}z_1z_2) \ .
  \eqn\eqz
$$
$g=0$ is an analytic equation in $z$ and always has five solutions.
Thus, $g=0$ is not equivalent to 
the original lens equation when the source is {\it outside} where 
there are only three images. 
However, in the  {\it inside} domains we are interested in here, 
the sources  result in the maximum number of images 5  
and so we can consider $0=g$ as the lens equation.  
If $a_j (\omega, \bar\omega); j =1,5$ are the zeros of $g$, 
we can factorize $g$ as  
$ g (z;\omega,\bar\omega) = a_{\circ}\Pi_{j=1}^5 (z-a_j)$, 
where $a_{\circ} = \bar\omega_1 \bar\omega_2$. \  
Now, we note from  \eqomegatwo\  that $J^{-1}$ is a very primary 
quantity we can calculate easily.  The value of $J^{-1}$ -- the amplification
combined with the parity -- at an image position $a_j$ is  
$$
  J^{-1} (a_j) = \partial_{\omega} z (a_j) = 
   - {\partial_\omega g\over\partial_z g}(a_j) 
   = \partial_\omega a_j (\omega, \bar\omega)  \ .  
 \eqn\eqJinv
$$
By summing it over all the images using \eqz, we find the ``sum rule"
completing the proof.  
$$
 \sum_{j=1}^5 J^{-1} (a_j) 
    = \partial_\omega \sum_{j=1}^5 a_j(\omega, \bar\omega) = 1  \ .  
\eqn\eqsumj
$$
The relevent terms in $g$  for the second equality are the highest order 
term for the whole polynomial and the next highest order term depending on 
$\omega$: $(z-\omega)z^4 \bar\omega_1\bar\omega_2$.  The constant 1 on the RHS
is determined by the relation between $\omega$ and $z$ in the factor
$(z - \omega)$.   

For a single lens, there are two images and $g$ is a quadratic polynomial
in $z$.  Since $g=0$ is saturated by the images (or $g=0$ is equivalent
to the lens equation), \eqsumj\ holds for a single lens.  Or, one can 
satisfy oneself by solving the quadratic lens equation and adding up 
$J^{-1}$ for the solutions. ($J = 1-|z|^{-4}$, where the lens is at $z =0$.)   

Now, the generalization of \eqsumj\  
for an arbitrary class of $n$-point lens systems
is only a matter of bookkeeping. 
The lens equation is given by 
$\omega = z - \sum_{j=1}^n \epsilon_j/\bar z_j \ ; \ \ 
               \sum_{j=1}^n \epsilon_j = 1 \ . $ 
We note that $f(\bar z)$ in \leqthree\ is a quotient
of a first order polynomial by a second order polynomial:
$f = f_1/f_2$.  For an $n$-point lens, $f(\bar z)$ is a quotient
of an $(n-1)$-th order polynomial by an $n$-th order polynomial:  
$f = f_{n-1}/f_{n}$. For a binary, the polynomial in \eqz\ \  
$g(z;\omega, \bar\omega) \equiv  g_2 \ 
  \supset  (z-\omega) \Pi_{j=1}^2(f_1 + \bar\omega_j f_2)  \ \supset 
 a_{\circ} (z-\omega) z^4$.  
 For an $n$-point lens,  the corresponding polynomial 
$g(z;\omega, \bar\omega) \equiv g_n   
  \ \supset  (z-\omega) \Pi_{j=1}^n(f_{n-1} + \bar\omega_j f_n)  
  \ \supset a_{\circ} (z-\omega) z^{n^2}$.  
Therefore, 
$$
  \sum_j  J^{-1}(a_j)
  = \partial_\omega \sum_j a_j (\omega, \bar\omega) = 1  \ .   \eqn\eqsumjn
$$
We  emphasize that the second equality of \eqsumjn\ holds only when 
the summation is over all the $n^2+1$ images, \ie, in the ``maximal
domains."  Only in the ``maximal domains", the lens equation is equivalent 
to $g=0$. 

\chapter{Graded Caustic Domains}

There is no ``inside a caustic" in a single lens ($n=1$) because
the caustic is point -- a ``collapsed caustic loop". 
For binary lenses ($n=2$), 
``inside a caustic" is ``a maximal domain" and the source plane consists
of one, two or three ``islands" of ``maximal domains" surrounded by
 the  {\it outside}.  When $n\ge3$, the caustics show more  
complicated structures as we have seen one of whose examples 
in figure \ternary.
Now, how do we sort these apparently complicated structures and understand
them intuitively?  Actually, it is rather simple to see that a source
plane consists of {\it graded caustic domains} -- namely, {\it domains}
with well-defined {\it degrees}.  We only need to  
recall a couple of well-known facts.   \ \  
First, the number of images of a source remains 
the same if the source does not cross a caustic curve.  So, we can spell out
the self-evident definition of a {\it domain}: 
{\it A domain of a source plane
is an area bounded by but devoid of caustic curves}. \  \   
Second, the number of images changes by two 
-- one positive and one negative --  
when a source crosses a caustic curve. 
Since caustic curves are oriented as determined by the phase angle of $\kappa$
we encountered in section 2, we can consider the sign of caustic crossings.
Then, given two randomly chosen domains, the algebraic sum of the number 
of caustic crossings with sign $\pm$ 
is independent of the paths connecting the two domains (see Rhie, 1995a for
more details). 
Now, define the {\it degree} of a caustic domain: 
The {\it outside} domain has {\it degree} 
zero and is denoted ${\cal D}^{\circ}$.  
{\it If the algebraic sum of the caustic crossing between a {\it domain} 
and {\it outside} is $m$, the {\it domain} has {\it degree}  $m$ 
and is denoted ${\cal D}^m$.}    (Lensing is a short range phenomenon,
and ${\cal D}^{\circ}$ can consist of many domains  where only one of them
includes infinity.)  Some of the  corollaries are : \ \  
1)\  ${\cal D}^{m}$ is always nested in ${\cal D}^{m-1}$'s 
and the number of images of a source in ${\cal D}^m$ is $n+1+2m$.  
\ \ 2)\  A {\it maximal domain} is where a source has  $n^2+1$ images
by definition.  Now, the  number of images of a source 
in ${\cal D}^{\circ}$ is $n+1$
and thus the {\it degree} of a {\it maximal domain} is $n(n-1)\over 2$.
If the maximum {\it degree} of the domains of an $n$-point lens is $N(n)$,
the source plane is  \ $\cup_{m=\circ}^{N(n)} \{{\cal D}^m\}$ where
$N(n) \le n(n-1)/2$.  We write ${\rm sup}N(n) = n(n-1)/2$.   

  \ Let's look at the example in figure \ternary:   
The source plane is made of one ${\cal D}^{\circ}$, nine
${\cal D}^1$'s, one ${\cal D}^2$ and one ${\cal D}^3$. 
As the seperation increases, the triple lens should converge to  
three single lenses: At a large separation, the source plane has three 
${\cal D}^1$'s near the lens positions.  The point we are getting at is  
that the fact that ${\cal D}^1$'s in the ``ocean" of ${\cal D}^{\circ}$ is 
a generic feature of the source plane of an arbitrary $n$-point lens 
when the ``correlations" of the lens masses is weak.  
As the masses  come close and 
their ``correlations" become strong, the ${\cal D}^1$'s merge and generate
higher hierarchical structures.  
In the case of binaries, however, ${\cal D}^1$ is also {\it maximal} and
merging does not result in higher degrees.

There is only one positive image for a source at infinity and thus for 
a source anywhere in ${\cal D}^{\circ}$.  
Therefore, the number of positive images of a source in ${\cal D}^m$
is $1+m$, and  the total amplification of the positive images is
$ A_+ ({\cal D}^m) \ge 1 + m$. 
A universal constraint on negative images is that they have 
non-vanishing amplifications unless the source is at infinity 
$\in {\cal D}^{\circ}$.  
Therefore,  $ A_{\rm tot}({\cal D}^m) > 1 + m$ when $m \ge 1$ 
(or ``inside the caustics").  
Inside a `maximal domain', the constraint \eqsumjn\  applies,
and $ A_{\rm tot} \ge n(n-1)+1$. Whether this inequality is saturated is
another question, and we will discuss this briefly below for $n\ge3$.

$ A_{\rm tot} = n(n-1)+1$ only if all the positive images are at the limit
points.  Since there are $2(n-1)$ (finite) limit points, 
$n$ has to satisfy that $2(n-1)=1+n(n-1)/2$.
In other words, $n=3$ is the only possibility besides the known cases of
$n = 1, 2$.   Indeed, we find that a source at the center of a symmetric 
ternary lens with  
separation (distance between two masses) 
$l = {\root 6 \of 2} \approx 1.12246$ has four 
positive images at the (finite) limit points.  Therefore, $A_+ = 4$
and $A_{\rm tot} = 7$. It should be an 
interesting question whether the solution we have found is unique or 
one of many.   For $n \ge 4$, the total amplification
of the maximum number of images is formally  $A_{\rm tot} > n^2+1-n $.   
It should be interesting to investigate the significance
of the high degree domains  in terms of the size and  frequency of 
the appearance in the source planes of multiple point lens systems.

Since this paper was submitted, the MACHO experiment has detected a binary
event toward the LMC (Bennett \etal, 1996, astro-ph/9606012) that shows 
the amplification
between two caustic crossings  less than 2. We have mentioned above
that $ A_{\rm tot} > 2$ inside any caustic of any $n$-point lens system.
Therefore, one can {\it definitely} conclude that the event was 
``contaminated" by other than geometric effect due to gravity (such as 
the third component of the lens). 
If the ``contamination" is due to blending of the source star that lies
in the bar of the LMC,   it may be verified by the centroid shift 
of the source star  (MACHO, private communication).

\singlespace
\ack
We thank D. Bennett for valuable input.  
This work was supported in part by
the U.S. Department of Energy at the Lawrence Livermore
National Laboratory under contract No. W-7405-Eng-48. 

\bigskip
\bigskip

\title {REFERENCES}

\def\aa{{\it A. \& A.}}
\def\apj{{\it ApJ}}

\def\ref{\par\hangindent=1cm\hangafter=1\noindent}
\parskip 0pt

\ref Alard, C., Mao, S., and Guibert, J., 1995,  submitted to \aa 

\ref Alcock, C., Akerlof, C.W., Allsman, R.A., Axelrod, T.S., Bennett, D.P.,
 Chan, S., Cook, K.H., Freeman, K.C., Griest, K., Marshall, S.L., Park, H.-S.,
 Perlmutter, S., Peterson, B.A., Pratt, M.R., Quinn, P.J., Rodgers, A.W.,
 Stubbs, C.W., and Sutherland, W., 1993, {\it Nature}, {\bf 365}, 621.

\ref Alcock, C., Allsman, R.A., Axelrod, T.S., Bennett, D.P.,
 Chan, S., Cook, K.H., Freeman, K.C., Griest, K., Marshall, S.L.,
 Perlmutter, S., Peterson, B.A., Pratt, M.R., Quinn, P.J., Rodgers, A.W.,
 Stubbs, C.W., and Sutherland, W., 1995, \apj, in press.

\ref Aubourg, E., Bareyre, P., Brehin, S., Gros, M., Lachieze-Rey, M.,
Laurent, B., Lesquoy, E., Magneville, C., Milsztajn, A., Moscosco, L.,
Queinnec, F., Rich, J., Spiro, M., Vigroux, L., Zylberajch, S., Ansari, R.,
Cavalier, F., Moniez, M., Beaulieu, J.-P., Ferlet, R., Grison, Ph.,
Vidal-Madjar, A., Guibert, J., Moreau, O., Tajahmady, F., Maurice, E.,
Prevot, L., and Gry, C., 1993, {\it Nature}, {\bf 365}, 623.

\ref Bennett, D.P. \etal,  AIP Conference Proceedings {\bf 336}: 
   Dark Matter,  eds., S. S. Holt and C. L. Bennett, 1995,  p.77.

\ref Bourassa, R. R., Kantowski, R, Norton, T. D., 1973, 
    \apj, {\bf 185}, 747. 

\ref Mao, S. and \pac, B., 1991, \apj, {\bf 374}, L37. 

\ref Rhie, S., 1995a,  preprint.  

\ref Rhie, S., 1995b,  preprint. 

\ref Tytler, D., \etal, 1995, in preparation.

\ref Udalski, A., Szymanski, M., Kaluzny, J., Kubiak, M., Krzeminski, W.,
Mateo, M., Preston, G.W., and \pac, B., 1993,
            {\sl Acta Astronomica} {\bf 43}, 289.

\ref Udalski, A., Szymanski, M., Kaluzny, J., Kubiak, M., Krzeminski, W.,
Mateo, M., Preston, G.W., and \pac, B., 1994, {\it Acta Astronomica},
{\bf 44}, 165.

\ref Witt, H.J., 1990, \aa, {\bf 236}, 311.

\ref Witt, H. J. and  Mao, S.,  1995, \apj, {\bf 447}, L105.

\endpage
\par \penalty-400 \vskip\chapterskip
   \spacecheck\referenceminspace \immediate\closeout\figurewrite
   \figureopenfalse
   \line{\fourteenrm\hfil FIGURE CAPTIONS\hfil}\vskip\headskip
   \input figures.texauxil
    
\endpage

\appendix

We would like to appreciate many anonymous referees  who have shown 
great interest  in this paper since it was submitted  a year ago. 
Since many of the issues are more general than is dealt with in  
this paper,  we would like to  include some of the discussions in this
appendix  starting with the ones more closely related to the content of 
this paper.  

{\bf 1}. \qquad   First, we show each step of the equations in (3.3)
starting with the expression of the inverse Jacobian matrix in
(2.2).  The first step is a simple manipulation of the implicit function
$g(z;\omega, \bar\omega)$ given in (3.2).   
From $0 = g(z;\omega, \bar\omega)$,  we get  
$$ 
 0 = dg(z;\omega, \bar\omega) = \del_z g dz + \del_{\omega} g d\omega
   + \del_{\bar\omega} g d\bar\omega  \ , 
$$
that is equivalent to 
$$
 dz = -{\del_{\omega} g\over \del_z g} d\omega  
      -{\del_{\bar\omega} g \over \del_z g} d\bar\omega \ .
$$
Therefore, we obtain the second equality of (3.3).  
$$
 \del_{\omega} z = - \del_{\omega} g / \del_z g  
$$ 
(This type of manipulations can be found in thermodynamics.  Thermodynamic  
functions are determined by any two of the thermodynamic variables 
-- density, pressure, temperature, entropy, etc --
and  the measurable quantities are  partial derivatives of the  
thermodynamic functions.)   One can calculate the coefficients 
(or the partial derivatives) $\del_{\omega} g $ and $\del_z g$ 
easily from the factorization of $g$, $g = a_{\circ}\Pi_{j=1}^5 (z-a_j)$
as given in the main text, where $a_j$ is a function of $\omega$
and $\bar\omega$,  and $a_{\circ} = \bar\omega_1 \bar\omega_2$ is
independent of $\omega$.   
$\del_{\omega} g (a_j) = a_{\circ} (-\del_{\omega}a_j)$  and 
$\del_z g (a_j) = a_{\circ}$  and one obtains the third equality.

{\bf 2}.  \qquad    We have shown that $A_{\rm tot} =3$
can be achieved inside a caustic of a binary lens only when two positive 
images have amplification $1$ and three negative images have amplification
$1/3$.   Having read through the proof, one can wonder, for example, why 
$A_1 = 1.2, \ A_2 = 1.1, \ A_3 = 0.2, \ A_4 = 0.3, \  A_5 = 0.2$ can not 
happen:  Can we understand it more intuitively?   
We believe what is stated in the main text 
explains fairly obviously why the positive images have to have $A=1$
each and the total amplification of the negative images is 1: \     
If $A_+$ and $A_-$ are the positive and negative total amplifications,
they satisfy the following relations  for $A_{\rm tot} = 3$.
$$
  A_+ + A_- = 3 \qquad ; \quad A_+ - A_- = 1 \ ,
$$
where the second equality is the sum rule
$\sum_{\rm images} J^{-1}  = 1$.
Therefore,  $A_+ = 2$ and $A_-=1$.   Since the amplification of positive
images are no less than 1, each positive image must have amplification
unity  --
in other words,  they are the images falling on the two limit points.
As is discussed in the main text,  the fact that the source position has to
be such that two positive images are at the limit points constrains the
source position to be at the mid-point between the lens positions and
the separation between the point lens elements to be $l = \sqrt{2}$.  The
class of two point lens systems is a two-parameter family given by
$\{\epsilon, l\}$.  The subclass of the two point lens systems which
allows the infimum amplification $A_{\rm tot} =3$  inside a caustic
is a one-parameter family given by $\{\epsilon\}$ because of the constraint
on the separation $l = \sqrt{2}$.   

This leaves us with the rest of the question  unanswered: \ 
Can we find  an intuitive (physical or structural)  
reason  why negative images all have the same  amplification $1/3$ for
$A_{\rm tot} =3$?  \  We are currently ignorant of an obvious reason, but
we  find it an interesting question to pursue.  That is especially so 
because a similar condition holds for the case of the triple lens 
we discussed in section 4:  The six negative images of the source with
$A_{\rm tot} = 7$ all have the same amplification  $1/2$.

{\bf 3}. \qquad  Here we present the calculations involved in the example
of the triple lens discussed in section 4, which follows exactly the same
line of analysis used for the case of binary lenses as described in the
main text.   The family of symmetric triple 
lenses is parameterized by the separation $l$.  If the center of mass is at
the origin, the lens positions can be given by 
$x_j = a \exp({i2\pi j\over 3}); \ j=0,1,2$, where $a = l/\sqrt{3}$.  
Then the symmetric triple  lens equation is
$$
  \omega =  z - {\bar z^2\over \bar z^3 - a^3} \ .   \eqn\Tleq
$$ 
The limit points are obtained by solving
$$
 0 = \kappa \equiv \del\bar\omega = {z^4 + 2a^3 z\over (z^3-a^3)^2} \ .
$$  
In other words, there are four limit points.
$$
 z_{\ast} = 0, \quad  {\root 3 \of 2}a e^{i\pi\over 3}, 
               \quad - {\root 3 \of 2}a,
               \quad  {\root 3 \of 2}a e^{i 5\pi\over 3} 
$$
Therefore, the source has to be at the origin, namely,
$$
 \omega_{\ast} = 0  \qquad \Leftarrow \qquad z_{\ast}=0  \ . 
$$ 
Now, in order for the other three limit points ($\not = 0$)
to be the image positions of the same source,  
the only free parameter $l$ (or $a$) has
to be fixed.   Because of the 3-fold symmetry of the lens system, 
we only need to consider one of them, say, the non-zero limit point
on the real axis.  From the lens equation \Tleq,  
$$
  0 = z_{\ast} - {z_{\ast}^2\over z_{\ast}^3 - a^3} 
  \qquad \Rightarrow \qquad    l^2 = {\root 3 \of 2} \ .  \eqn\Tleqtwo 
$$
Therefore,  the symmetric triple lens ($\epsilon_j = 1/3; j =1,2,3$) 
with separation $l = {\root 6 \of 2} \approx 1.12246$ accommodates
a source position where  all the positive images fall on the limit points 
(amplification unity).  Applying the constraint of the sum rule in the 
maximal domain ${\cal D}^3$,   one finds that the total amplification  
of the source at the origin is $A_{\rm tot} = 7$ \ ($A_- = 3$) as is 
stated in the main text.   Incidentally, the non-zero limit point on
the real axis is $z_{\ast} = - {\root 3 \of 2}a = - \sqrt{2/3}$.  

We might as well spell out the positions and amplifications of the  
negative images, which is a straightforward exercise one can carry out
analytically.  We note that the real lens equation \Tleqtwo\  has 
four real solutions.   Two are the ones on the limit points we discussed
above.   The other two solutions satisfy  the second order polynomial
equation.  
$$
  0 = z^2 - \sqrt{2\over 3} z - {1\over 3}  \qquad \Rightarrow \qquad 
  z_{\pm} = {1\over \sqrt{6}} \pm {1\over \sqrt{2}}   \eqn\Tleqthr
$$ 
Now, we can show that the Jacobian determinant at $z_{\pm}$ is
$J_{\pm} = - 2$.   We recall that
$$
 J = 1 - |\kappa|^2 \ ; \qquad 
  \kappa \equiv \del\bar\omega = {z^4 + 2 a^3 z \over (z^3-a^3)^2} \ .
$$
Using the first equation in \Tleqthr,  one can reduce 
$\kappa$ into a quotient of two linear order polynomials.
$$
 \kappa = {6 a^3 z + 1/3 \over 3 a^3 z + 1/3} 
$$
Using the same reduction method,  one can find that $\kappa^2 = 3$
and hence $J = -2$. 
Up to now, we have discussed only the negative images on the real axis. 
The other four negative image positions can be found by rotating these 
two solutions on the real axis by $2\pi/3$ and $4\pi/3$ with respect to
the origin.  In other words, the six negative images of the source at 
the center of mass are located at
$$
  z = z_{\pm}, \quad z_{\pm} e^{i2\pi\over 3}, 
               \quad  z_{\pm} e^{i4\pi\over 3}   \ .
$$ 
Now, the symmetry of the lens system guarantees that all the negative
images have $J = -2$ (and hence $A = 1/2$). 

{\bf 4}.  \qquad   The  lens equation in complex coordinates
is accepted well in the community, but the differentiation of the
equation is not.   However,  the Jacobian determinant 
is one of the most important  observable quantities  in gravitational lensing,
and one should know how to calculate it in whatever coordinate system one
has chosen.   We have chosen complex coordinates and hence it is most 
natural to examine the linear behavior of the lens equation in the same
complex coordinates.   The necessity to calculate the Jacobian determinant
arises from that the relative brightness of an image of a lensed star is
given by the ratio of the sizes of the image and its source star.
In other words,  we need to know what happens to an infinitesimal area
element in the lens plane under the mapping given by the lens equation.
(The lens equation is an explicit mapping from an image position to its
source position.)  

If we parameterize the lens plane by Cartesian 
coordinates $(x,y)$, the infinitesimal area element is given by
$dx\wedge dy$, where $\wedge$ stands for the exterior product (or ``cross
product"): \ $dx\wedge dy = - dy\wedge dx$.   (One must be familiar 
with ``cross product" from fluid mechanics or electrodynamics.) 
In terms of the complex coordinates $(z, \bar z)$,  the area element 
(always real) is given by a multiple of $dz\wedge d\bar z$.   
We can calculate the multiplicity
using  $z = x + iy$ and $\bar z = x - iy$.   
$$
   dz\wedge d\bar z = (dx+i dy)\wedge (dx-i dy)
                    = dx\wedge (-i dy) +  i dy\wedge dx
                    = -2i dx\wedge dy \ .  
$$
Therefore, the multiplicity is $i/2$.    
Now, in order to compare the sizes of an infinitesimal image and its
source,  it suffices to calculate  the ratio
$d\omega\wedge d\bar\omega / dz\wedge d\bar z$  
because the multiplicity $i/2$ is a constant.  
$$
   d\omega\wedge d\bar\omega
 = (\del\omega dz + \bar\del\omega d\bar z)\wedge
    (\del\bar\omega dz + \bar\del\bar\omega d\bar z)
 = (\del\omega \bar\del\bar\omega - \bar\del\omega \del\bar\omega )
    dz\wedge d\bar z  \ . 
$$
Therefore, 
$$
  {d\omega\wedge d\bar\omega \over dz\wedge d\bar z} 
 = |\del\omega|^2 - |\bar\del\omega|^2  \ .  \eqn\APone
$$
It is well known in the community that the RHS is nothing but the 
Jacobian determinant  $J = |\del\omega|^2 - |\bar\del\omega|^2$.  
One can get this expression of $J$ in complex coordinates from the
familiar expression of $J$ in the real Cartesian coordinates  
by replacing the coordinates by complex coordinates
using $z=x+iy$ and $\bar z = x - i y$.     

Jacobian determinant literally means determinant of Jacobian matrix. 
Where does the Jacobian matrix come into picture in the landscape of
lensing?    We recall that we need to know what happens to its source
when an image changes its position slightly (or infinitesimally).  
In other words,  we need to know the linear behavior of the lens equation.
$$
 \eqalign{d\omega &= \del\omega dz + \bar\del\omega d\bar z  \cr
          d\bar\omega &=
               \del\bar\omega dz + \bar\del\bar\omega d\bar z \cr} \ .
  \eqn\APtwo
$$ 
These coupled linear equations can be organized as follows.  
$$
 {d\omega\choose d\bar\omega}
  = \pmatrix{\del\omega & \bar\del\omega \cr
             \del\bar\omega & \bar\del\bar\omega \cr}
    {dz\choose d\bar z}  \ ,    \eqn\APthree
$$
where the usual matrix multiplication convention applies.
This ($2\times 2$) matrix is called Jacobian matrix (to honor the
historical development), and one can see that the determinant of the
matrix gives rise to the expression of $J$ in \APone.

The equation \APthree\  shows clearly that $\del_z\omega$ and
$\del_{\bar z}\omega$ in (2.2) and their complex conjugates
constitute the Jacobian matrix elements in complex coordinate basis.
However, this could mean a dilemma if one tries to recover the familiar
Jacobian matrix in Cartesian coordinate basis by replacing the complex
variables by the Cartesian coordinate variables because one is sure to fail! 
We mentioned above that exactly the same process works for the Jacobian
determinant.  So, what is wrong?    Actually,  there is nothing wrong.
It's simply that Jacobian matrix ``rotates"  accordingly with the change
of coordinate basis while Jacobian determinant remains invariant.
Let's consider changing coordinates
where the new coordinate basis is obtained by multiplying the old 
coordinate basis by a matrix $V$ from the left.  For example, 
$$
  V {d\omega\choose d\bar\omega} \Leftarrow {d\omega\choose d\bar\omega} 
  \ ;  \qquad  V {dz\choose d\bar z}  \Leftarrow {dz\choose d\bar z} \ .
$$
Now,  we can rewrite \APthree\ in terms of the new basis.
$$
 V {d\omega\choose d\bar\omega}
  = V \pmatrix{\del\omega & \bar\del\omega \cr
             \del\bar\omega & \bar\del\bar\omega \cr} V^{-1} \ \  
    V {dz\choose d\bar z}      \eqn\APfour
$$  
In other words, the Jacobian matrix ``rotates" through  a similarity
transformation ${\cal J} \mapsto V {\cal J} V^{-1}$.
If one tries to understand the Jacobian matrix in complex coordinates
starting from the more familiar form in Cartesian coordinates, 
one should make sure to rotate the matrix as well as change the coordinate
functions.   Without the proper rotation carried out, 
one can erroneouly conclude that the  matrix elements in (2.2) are wrong. 
On the other hand,  the determinant of a matrix is invariant under a
similarity transformation unlike the Jacobian matrix.  
In other words, the Jacobian determinant
is a scalar and is coordinate independent.  That shouldn't be surprising
because the Jacobian determinant is an observable quantity.   (Coordinate
dependence of the Jacobian determinant would mean that the  amplification 
of a star would change  depending on which coordiante  system we use to
calculate it.  That is undesirable.)  

We might as well discuss why the inverse Jacobian matrix elements
are given by (2.3).   In order to take advantage of the Einstein summation
convention,  let's  rename the complex variables as follows: \
$X_1 = z, X_2 = \bar z$ and $Y_1 = \omega, Y_2 = \bar\omega$.
Then the Jacobian matrix elements of the lens equation are
$\partial Y_c / \partial X_a; a, c = 1, 2$.   And the inverse Jacobian
matrix elements are given by $\partial X_b / \partial Y_c; c, b = 1, 2$
as one can easily check using the product rule
$ \sum_c {\partial Y_c\over \partial X_a}
          {\partial X_b\over \partial Y_c} = \delta_{ab} $.
In other words, $\del_{\omega}z$ and $\del_{\bar\omega}z$ in (2.3)
and their complex conjugates constitute the inverse Jacobian matrix
elements. (In case one is not familiar with the product rule,  
see the formula above (3.2.11) in S. Weinberg, 1972, 
``{\it Gravitation and Cosmology}", John Wiley \& Sons, Inc.)

{\bf 5}. \qquad  There have been some concern whether it is proper
to denote the Jacobian determinant by $J$ or something like det$J$
should be used where $J$ denotes the Jacobian matrix.  In our opinion,
as is reflected in our literary work here, the Jacobian determinant
deserves a simple notation just as much as the amplification deserves
a widely accepted simple notation $A$ or $\mu$.   Jacobian 
determinant is a determinant mathematically and one can prefer a more
explicit notation such as det$J$.   However, the importance of the 
determinant in lensing lies in that the determinant is a measurable
quantity.  The magnitude of the determinant is nothing but $1/A$ and
the sign of the determinant is the parity of the image.  Furthermore, 
the determinant is a very important quantity in the understanding of the 
lens plane.  In other words, the determinant $J$ plays a far more
important role in lensing than is implied by the mere fact that the
function is determinant of a matrix.  As a minor comment, this is not
the first occasion where the determinant is denoted by $J$.  See
Kayser \etal, 1986, A\&A, {\bf 166}, 36. 
 (This paper contains a mistake  in the definition of $J$, but
it is an easily recognizable  mistake one can not be confused  by.) 
On the other hand,  we are not insisting that everyone in the lensing
community should use the same notations as we do.  That is because
physics lies in the structure of the system indenpedently of which letter 
of which alphabet system is  used for which variable.

{\bf 6}. \qquad  We would like to direct the attention to the last 
paragraph in section 2.  The original purpose of the example discussed
in this paragraph is to show how the sum rule fails in a ``non-maximal
domain".    However, one can also examine the behavior of the images
at a cusp crossing.   The merit of this example is that it is analytically
solvable leaving no room for numerical artifacts and ambiguities.
 As the caustic bifurcates when $l=2$, two images disappear. 
When $l \rightarrow 2$ and the source nears the cusp, the amplification 
of these two images increases as $1/\sqrt{a}$ where $a$
is the distance of the source from the cusp.  These are all expected.
However, what may not have been expected is that {\it the disappearing 
images are  both positive.}  Isn't it contradictory to our conventional
wisdom that {\it two images of opposite parity disappear (or appear)
at a caustic crossing}?   Strictly speaking, yes.   
At a cusp crossing, the images that disappear together have the same parity,
while the two disappearing images at a line caustic crossing have the
opposite parities.   On the other hand, at a cusp crossing, there is the 
third image that crosses the critical curve where the two images 
disappear.   In other words, the third image changes the parity (from 
negative to positive here), and there is no net change of the total 
parity of the images.   This third image is the one
whose amplification increases as $1/a$ as the source nears the cusp
similarly to the images of a single lens.  Therefore,  the conventional
wisdom should be modified to the correct theorem that 
{\it the number of images change by two -- or a multiple of two where
the caustics self-intersect -- and the total parity remains the same 
at a caustic crossing}.     

One can find an illustration of cusp crossing in figure 8  
in Schneider and Weiss, 1986, A\&A 164, 237.  
(The curvatures of the critical curve and that of the ``off-axis"
 image trajectory in figure 8 have the opposite signs, but they can also
have the same sign depending on the lens parameters.   At a cusp crossing
on the lens axis, they always have the same sign.)

{\bf 7}. \qquad  One of the most interesting issues seems to be  
the controvery over our identification of lensing as  the inverse
mapping of the lens equation (the bottom of page 3).

Let's consider the {\it lensing} of a star by a black hole.
The BH dictates the geometry of the space around it, and the light
from the star to an observer (or us) follows the null geodesics (or 
optical paths) allowed by the geometry of the ``curved" space.  
{\it Lensing} refers to {\it the phenomenon} that the starlight follows the 
null geodesics of the curved space instead of the optical path the light
would follow were there not be the BH.    Each null geodesic ``delivers"
light from the star to the observer in its own way,  and the light arrived
at the observer through a specific optical path is called an image.  
The observer can see as many images as the number of null geodesics 
between the target star and the observer. 
Therefore, {\it lensing} refers to  {\it the process forward in time},
namely, the propagation of the light from the light source through the 
available null geodesics to corresponding images, 
and thus it is sensible to talk  about {\it lensing of a star}.    
On the other hand,  propagation of light along a given null geodesic 
is a time reversible process.  For example, the observer can send the light
back to the source star by putting a mirror perpendicular to the optical
path at the position of the image. (Orthogonality is well defined in 
general relativity.)   In other words, the process {\it lensing}
has a well defined {\it inverse process}.   

One can study the optical path lengths through time delay 
measurements.  One can study the spin of the photons or collectively
the polarization of the light in the images.  Or, one can study the  
beams of the photons -- propagation vectors (or angular positions)
and sizes -- arriving at the observer.  The lens equation addresses
the relation of the angular positions of the images and their source star.   
More specifically,  the lens equation  determines the position 
of the source star for a given image position.  In other words, the 
lens equation is an {\it explicit mapping} describing the {\it inverse
process}.    

When we introduce a model (here the lens equation), the physical process
(lensing here)  must be {\it identified} as a mathematical process within 
the mathematical system.   (If not, what is the use of the model?)    
So, what is lensing as a mathematical process within the mathematical
system defined  by the lens equation?   The question boils down to 
whether the {\it inverse process} has well defined inverse.  
The lens equation is a {\it projection map} from the lens plane 
onto itself where three or five images are mapped into one source position.  
Therefore, the inverse of the projection map ({\it inverse process})
is the mapping from the lens plane to itself where a source position 
is mapped into multiple images.  This  {\it multiplicity} of the images 
requires more specific definition of the {\it inverse}: \ Given an image, 
the restricted mapping from the image to its source and back to the same 
image is a well defiend identity mapping.  In other words, the lens equation 
is invertible when restricted to one given image.  
This type of invertibility is said to be {\it locally invertible}.  
In other words, {\it the lens equation is locally invertible}.   
But, not everywhere.  The mapping 
can be restricted to one image only when the image is distinguishable
from the other images of the same source.  Where two images merge to 
disappear or split to appear -- in other words, on the critical curve,
the distinguishability fails.  This singular behavior is manifested
in the factor $1/J$ in the inverse Jacobian matrix elements in (2.3).
Therefore, the inversion property of the lens equation is correctly
stated as follows. {\it The lens equation is locally invertible except
on the critical curve.}   We should note that the physically 
measurable quantities such as magnification and distortion of the images
belong to the local behavior.

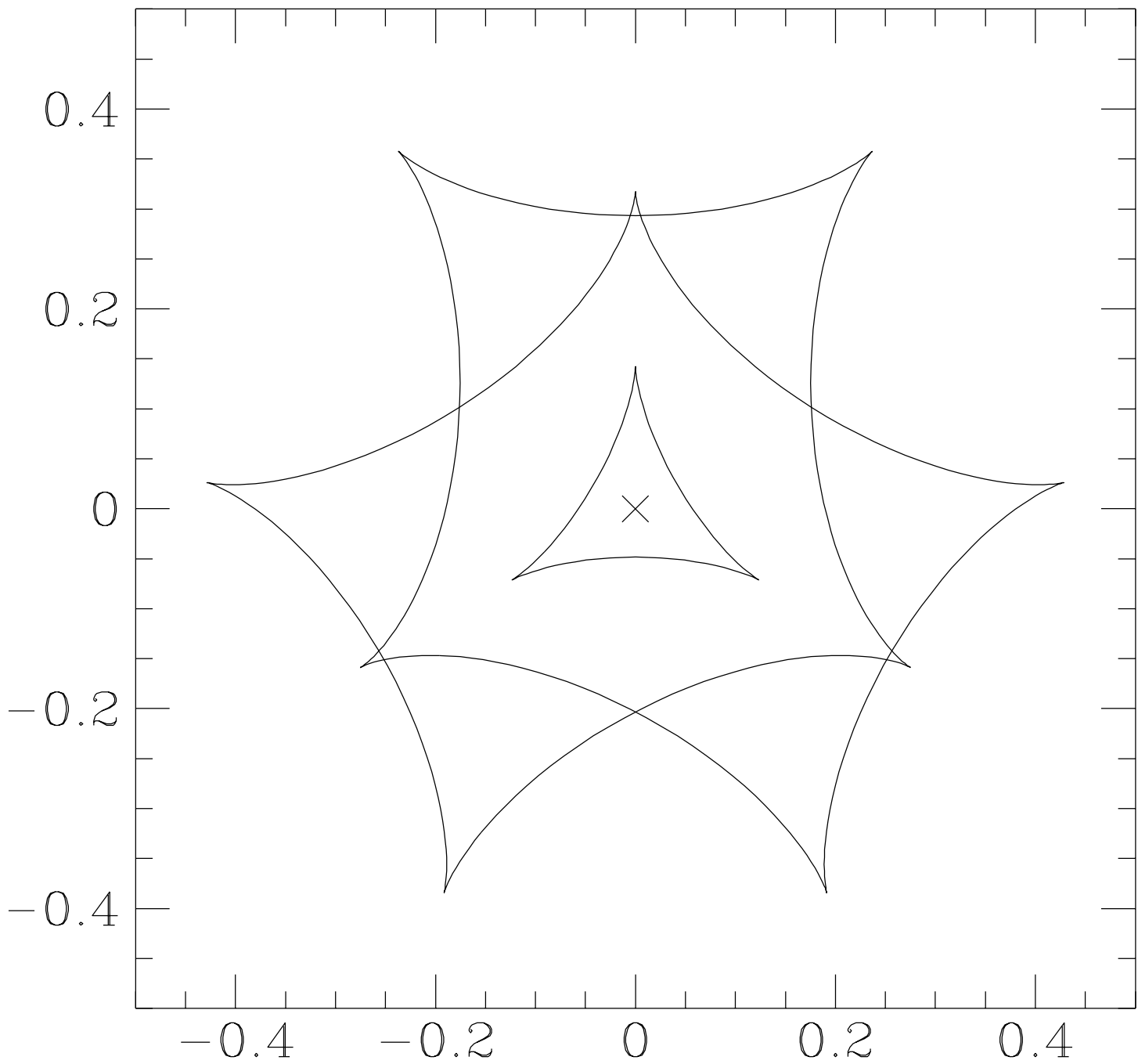
\end